\documentclass[useAMS,usenatbib]{mn2e}
\usepackage{times,graphicx,natbib}
\newcommand{\apj}{ApJ}                 \newcommand{\apjs}{ApJS}
\newcommand{\mnras}{MNRAS}             \newcommand{\aap}{A\&A}
             
\newcommand{\aj}{AJ}                   \newcommand{\na}{New Astron.}

\newcommand{\physrep}{Phys.~Rep.} \newcommand{\dd}{{\rm d}}
\title[Complementary second law of thermodynamics]
{Entropy principle and complementary second law of thermodynamics for self-gravitating systems}

\author[P. He and D.-B. Kang]{Ping He$^{1, 2}$\thanks{Email: hep@itp.ac.cn} and
Dong-Biao Kang$^{1, 2}$ \\
$^1$ Key Laboratory of Frontiers in Theoretical Physics, Institute of Theoretical Physics, Chinese Academy of Sciences, Beijing 100190, China \\
$^2$ Kavli Institute for Theoretical Physics China, Chinese Academy of Sciences, Beijing 100190, China}
\date{\today}
\begin{document}
\label{firstpage} \maketitle

\begin{abstract}
The statistical mechanics of isolated collisionless self-gravitating systems is a long-held puzzle, which has not been successfully established for nearly 50 yr. In this work, we employ a phenomenological entropy form of ideal gas, first proposed by White \& Narayan, to revisit this issue. By calculating the first-order variation of the entropy, subject to the usual mass- and energy-conservation constraints, we obtain an entropy stationary equation. Incorporated with the Jeans equation, and by specifying some functional form for the anisotropy parameter $\beta$, we numerically solve the two equations, and demonstrate that the velocity anisotropy parameter plays an important role to attain a density profile that is finite in mass, energy, and spatial extent. If incorporated again with some empirical density profile from simulations, our theoretical predictions of the anisotropy parameter, and the radial pseudo-phase-space density $\rho/\sigma^3_r$ in the outer non-gravitationally degenerate region of the dark matter halo, agree rather well with the simulation data, and the predictions are also acceptable in the middle weak-degenerate region of the dark halo. The disagreements occur just in the inner strong-degenerate region because of the ignoring of gravitational degeneracy. As far as we know, our results may be the first theoretical predictions based on the entropy principle that can partially match the empirical data. The second-order variational calculus reveals the seemingly paradoxical but actually complementary consequence that the equilibrium state of self-gravitating systems is the global minimum entropy state for the whole system under long-range violent relaxation, but simultaneously the local maximum entropy state for every and any small part of the system under short-range two-body relaxation and Landau damping. This minimum-maximum entropy duality means that the standard second law of thermodynamics needs to be re-expressed or generalized for self-gravitating systems. We believe that our findings, especially the complementary second law of thermodynamics, may provide crucial clues to the development of the statistical physics of self-gravitating systems as well as other long-range interaction systems.

\end{abstract}


\section{Introduction}
\label{sec:intro}

Cosmological $N$-body simulations of cold dark matter have revealed many empirical relationships concerning collisionless self-gravitating dark matter haloes, such as the density profile, the pseudo-phase-space density profile, the distributions of shape and spin parameters, the velocity-anisotropy-parameter profile and the distribution of velocity dispersion \citep{navarro08, white09}. They used to be considered as being universal, i.e. different cold dark matter haloes obey the same empirical laws only with different scale parameters, which are independent of the halo mass, the initial density fluctuation spectrum and the values of the cosmological parameters. The famous one of these universalities is the universal density profile of cold dark matter haloes. These universal profiles are usually expressed as NFW \citep*{nfw95, nfw96, nfw97, bullock01} or \citet{moore99} or even other specific forms, such as the \citet{einasto65} profile. We should notice that, however, more and more recent high-resolution $N$-body simulations claim that the dark halo density profiles are not strictly universal, which may depend on halo mass, merger history or substructures \citep*[see][]{jing00, subramanian00, klypin01, fukushige04, zhang04, simon05, merritt06, ricotti07, navarro08}.

In fact, long before these simulation results of dark matter haloes were found, astronomical observations had already revealed that some self-bounded gravitating systems exhibit remarkable structural regularities. For example, elliptical galaxies display a quasi-universal luminosity profile described by \citet{dev48}'s $R^{1/4}$ law, and globular clusters are generally well fitted by the Michie-King model
\citep{galdyn08}.

Although these empirical relationships are simple and practical, we just know what, but not why they behave in such ways, since they have no physical basis. Up to now, however, few of the physical origins concerning these empirical results have been investigated, except for the universal dark halo density profiles, whose origin was explored as soon as the universality was found \citep{white98}. Thereafter, many other explanations for the origin of dark halo density profiles were proposed, and most of them, similar to the one by \citet{white98}, are concentrated on various dynamical processes, such as collapsing, mergers of subhaloes or accretion of surrounding materials \citep*{huss99, nusser99, ma00, taylor01, manrique03, ascasibar04, williams04, lu06, bellovary08, elzant08}. However, a recent work by \citet{white09} suspected this kind of dynamical origins. According to their simulations of hot dark matter haloes, they claimed that mergers do not play a pivotal role in establishing the universalities, and thus their results contradict the models which explain the universalities as consequences of mergers.

Is it possible that all these empirical relationships originate from statistical mechanics, or more concretely, do they reflect the properties of self-gravitating systems in statistical equilibrium states? If so, from theories of thermodynamics and statistical mechanics, we know that an equilibrium state of a thermodynamic system does not depend on any specific dynamical processes or mechanisms of how to attain it. If dark matter haloes really attain the statistical equilibrium states, then it would be very easy to understand the finding by \citet{white09}, why the universalities are irrelevant to mergers.

The investigation of equilibrium statistical mechanics of isolated collisionless self-gravitating systems is a long-held issue, which can date back to the early 1960s. \citet{ant62} proved that global entropy maxima do not exist for self-gravitating systems, which might be the first study that exhibits the great complexity of the statistical mechanics for self-gravitating systems. Later, \citet{lb67} in his seminal work proposed `violent relaxation' to account for the formation process of elliptical galaxies. Just in the same work, he derived a `most probable' distribution of a system in its `equilibrium state' that has infinite mass, energy and spatial extent, and hence is not realistic and not acceptable. Following \citet{lb67}'s work, \citet{shu78}, \citet*{kull97} and \citet{nakamura00}, with increasing complexity of phase-space partitions, reworked for their own coarse-grained distribution schemes. These authors aimed to remedy the flaws in the previous works, the most significant of which is the mass segregation problem, i.e. in a multi-species composition, the heaviest objects tend to concentrate more centrally whereas the lightest ones tend to go towards the border. However, these authors did not overcome the problem of infinite mass that arises from \citet{lb67}'s statistical approach. We refer the reader to \citet{bindoni08} for the comprehensive review about the history of these works.

Since the concept of entropy is the basis of statistical mechanics, ever since \citet{ant62}, who used the usual Boltzmann-Gibbs entropy, various alternative entropy forms have been attempted to account for the stability and equilibrium configurations of self-gravitating systems, such as the generalized $H$-function \citep*{tremaine86}, the entropy of ideal gas \citep{white87} or Tsallis entropy \citep{tsallis88, hansen05}. There are also other authors using entropy arguments to draw conclusions about the final states of violent relaxation \citep[e.g.][]{stiavelli87, spergel92}. All these entropies are maximized, subject to the usual mass- and energy-conservation constraints \citep{ant62, lb67, tremaine86, white87}, or other additional conditions, such as hydrostatic equilibrium \citep{white87}.

Usually, the equilibrium state of a thermodynamic system is defined by employing the principle of maximum entropy, which states as follows: the realistic thermodynamic relations of the stable equilibrium state for any thermodynamic system may be derived by seeking the probability density function that maximizes the Boltzmann-Gibbs entropy (or other plausible entropy forms), subject to all relevant constraints. However, the principle of maximum entropy for self-gravitating systems has been continually questioned for the past several decades. For example, there is a famous argument, originated by J. Binney, against the feasibility of Boltzmann-Gibbs entropy, in that such an entropy form is unbounded above (\citealt{galdyn08}; see also \citealt{tremaine86, white87}), which resembles \citet{ant62}'s earlier result that there exists no global maximum entropy for self-gravitating systems. Confirming an earlier study by \cite{sridhar87}, \citet{soker96} showed that the general $H$-functions during the process of violent relaxation, may not necessarily increase at all times and may decrease if the system has specific, far from equilibrium, conditions. They further proposed to distinguish between two phases of violent relaxation, i.e. in one phase, the $H$-function may decrease with time, whereas in the other it is always a non-decreasing function of time. This result suggests that whether $H$-function is a monotonically increasing function of time during the violent relaxation process is still an open question \citep{kandrup87}, or equivalently, the equilibrium state of self-gravitating systems should not be simply considered as the maximum entropy state \citep{chavanis02}.

Worse still, \citet{lb05} revealed that there exists an inconsistency in theories of violent relaxation with \citet{lb67} and \citet{nakamura00}'s distributions. The inconsistency arises from the non-transitivity of these theories, i.e. a system that undergoes a violent relaxation, then followed by another, with an addition of energy for the latter, is not equivalent to the one that undergoes the whole process directly from the initial to the final state. The non-transitivity is destructive, since the `equilibrium state' is dependent on the evolutionary path of the system, which implies that a complete methodology of statistical mechanics is probably impossible for self-gravitating systems, and should be replaced by some dynamical approaches.

Therefore, after unsuccessful attempts for nearly 50 yr, the concept of entropy as well as the methodology of statistical mechanics might have been widely regarded as inviable for self-gravitating systems. As a result, we are in such an embarrassing situation that we have few applicable analytic theories concerning self-gravitating systems, such as elliptical galaxies and dark matter haloes, except for the Poisson-Vlasov equations (or the Jeans equation) and the fitting formulae from either $N$-body simulations or observations.

At the same time, these problems and difficulties agonize statistical physicists as well. They raised such questions for the general long-range interaction systems, as non-additivity or non-extension, ensemble inequivalence and negative specific heat in microcanonical ensemble and ergodicity breaking \citep{campa09}. Moreover, since it is not known how to define the equilibrium state of a self-gravitating system, such a state that an equilibrated system attains is vaguely called the quasi-stationary state.

It seems that, in order to establish a successful statistical mechanics of self-gravitating systems, some new ideas or even conceptual revolutions may be desired to get out of this long-term impasse. In this work, we explore whether the puzzle concerning the equilibrium statistical mechanics of self-gravitating systems can be solved by introducing some new thoughts, or by revising some fundamental laws or principles of physics.

Our paper is arranged as follows. In Section~\ref{sec:basic}, we firstly propose a mean-field model to separate gravitational forces into long- and short-range parts, and subsequently identify three distinct relaxation mechanisms, i.e. violent relaxation, two-body relaxation and gravitational Landau damping. Then by calculating the first-order variation of the entropy of \citet{white87}, we derive an entropy stationary equation. In Section~\ref{sec:result}, combined with the Jeans equation, we solve the two equations in two cases. First, we show that the velocity anisotropy parameter is of great significance to make the density profiles of dark matter haloes have finite mass, energy and extent. Then, ignoring the gravitational degeneracy in the inner region of the dark matter halo, we find that the predicted anisotropy parameter and the pseudo-phase-space density agree rather well with the simulations of \cite{navarro08} in the outer non-degenerate region, and are also in rough agreements with the simulations in the middle weak-degenerate region of the dark halo. An upturn deviation from the power law of the pseudo-phase-space density at the halo outskirts is also predicted, which is consistent with the finding by \citet{ludlow10}. In Section~\ref{sec:gen2nd}, we generalize the standard second law of thermodynamics to long-range self-gravitating systems, which states in the complementary way as follows: for an isolated self-gravitating system, the entropy can {\em automatically decrease} under long-range violent relaxation, until violent relaxation terminates and the system attains the equilibrium state with the global {\em minimum} entropy for the whole system; yet, in this very equilibrium state, the system is simultaneously in the local {\em maximum} entropy state for every and any part of the system under short-range two-body relaxation and Landau damping. In Section~\ref{sec:cons}, we conclude that, with this revised, complementary second law of thermodynamics, the concept of entropy is still valid, and thus in principle, the equilibrium statistical mechanics can also be established for such long-range self-gravitating systems.

\begin{figure}
\includegraphics[width=\columnwidth]{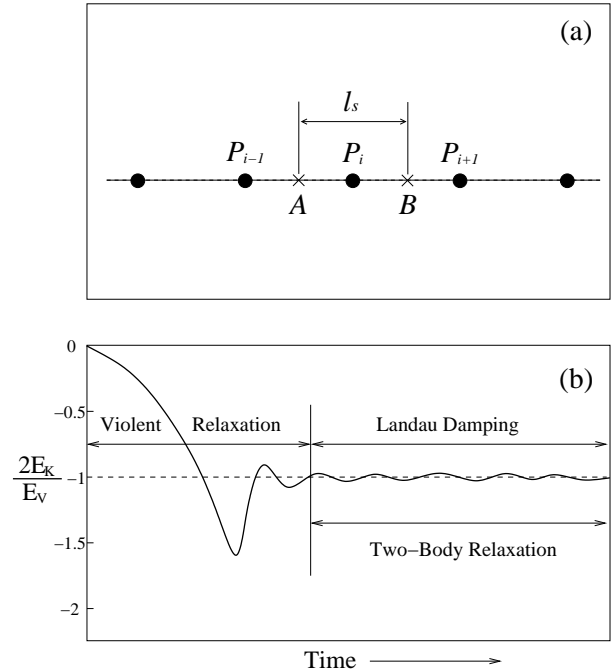}
\caption{(a) Illustration of the separation of gravitational force. The system possesses some geometrical symmetry in the equilibrium state, so that the long-range parts of gravitational forces of the particles can be collectively manifested as a mean-field force, with short-range residual effects determined by nearby particles. (b) The three relaxation mechanisms of self-gravitating systems, i.e. violent relaxation, two-body relaxation and Landau damping. The system is driven by violent relaxation into the virial equilibrium state, which is maintained by Landau damping and two-body relaxation.}
\label{fig:fig1}
\end{figure}

\section{Basic Theories and Formulae}
\label{sec:basic}

\subsection{Mean-field model for self-gravitating systems}
\label{sec:mfa}

It is usually believed that, undergoing the violent relaxation procedure \citep{lb67}, a self-gravitating system can attain a so-called quasi-stationary state \citep{campa09}, i.e. the virial equilibrium state. Throughout this paper, we tentatively call this quasi-stationary state as equilibrium state.

The appropriate gravitational model should be the mean-field model, which is the first step towards the correct statistical mechanics of self-gravitating systems. The following is a brief introduction to the idea of mean-field model, which was suggested by \citet{labini08}. For the detailed discussion of the mean-field model for self-gravitating systems, we refer the reader to their original article. There are also many discussions about the mean-field model that can be found in \citet{galdyn08}.

Usually, the standard statistical mechanics deals with short-range interaction systems. The typical feature of these systems is that the interaction between two particles is strong and repulsive when they are very close to each other, but on large distances no force is exerted between two particles. As a result, particles most of time move at nearly constant velocity, and then are subject to violent and instantaneous accelerations when they collide with one another. In this situation, the typical relaxation mechanism arises from two-body collisions.

One of the difficulties of developing the statistical mechanics of self-gravitating systems lies in that gravitational interaction is long-range attractive force, which means that a particle is coupled to almost all other particles at all spatial scales of the system. So it seems that all different range of scales are equally important to the dynamical properties of self-gravitating systems. Fortunately, we can circumvent this obstacle by constructing a gravitational model, in which the long-range part of gravitational force between the particles is collectively manifested as a mean field for a finite self-gravitating system, and the short-range part arising from the nearby particles is treated as fluctuations in this mean gravitational field. Note that the equilibrated self-gravitating systems usually possess some geometrical symmetry, by virtue of which the long-range coupling between the particles can be broken when the system is in its equilibrium states.

We illustrate this separation of gravitational force with a simplified one-dimensional model as follows. See Fig.~\ref{fig:fig1}(a), for the one-dimensional system in equilibrium, the particles are homogeneously distributed in the mean field, indicated by the dashed line, with a mean separation as $l_s$. In this model, if a particle is initially located at point $A$, it will not feel any gravitation. When the particle moves towards the right side, however, it will be more and more affected by the gravitational force of particle $P_i$. When the particle crosses over $P_i$ on the right side, continuously moving farther and farther away, the gravitational force exerted on this particle becomes less and less with the increasing distance from $P_i$, and vanishes again at point $B$. So from this model, we can see that after removing the mean-field force, the effective short-range part of gravitational force of a particle is just dictated by its nearby particles. We separate the gravitational potential and the density as $\Phi({\textbf{\emph x}},t) = \Phi_b({\textbf{\emph x}},t)+\delta\Phi({\textbf{\emph x}},t)$, and $\rho({\textbf{\emph x}}, t)=\rho_b({\textbf{\emph x}}, t) + \delta \rho({\textbf{\emph x}}, t)$, in which $\Phi_b({\textbf{\emph x}},t)$, $\rho_b({\textbf{\emph x}})$ are the long-range mean-field quantities, and $\delta\Phi({\textbf{\emph x}},t)$, $\delta\rho({\textbf{\emph x}}, t)$ are the short-range residuals, respectively. With Poisson equation $\nabla^2\Phi = 4{\pi}G\rho$, these long- and short-range dynamical variables satisfy their respective Poisson equations
\begin{eqnarray}
\label{eq:poisson}
    \nabla^2\Phi_b = 4{\pi}G\rho_b,  \nonumber \\
\nabla^2\delta\Phi = 4{\pi}G\delta\rho.
\end{eqnarray}
Thus, we can see that, in this mean-field model, the mean, smoothed-out gravitational potential is treated as a background field, corresponding to large-scale perturbations; whereas the short-range residual effects are regarded as fluctuations in this background, corresponding to small-scale perturbations. In the mean-field model, a particle couples not directly with particles at large separations, but only with this background potential, with the residual short-range effects determined by a few particles nearby. In later sections, we will see the great significance of this separation of gravitational force, as well as different physical implications by distinguishing between these two types of perturbations.

\subsection{Violent relaxation, two-body relaxation and Landau damping}
\label{sec:relax}

It is the instability of the mean gravitational potential, i.e. $\partial \Phi_b / \partial t \neq 0$, that induces a strong, long-range relaxation mechanism, called violent relaxation \citep{lb67}. When violent relaxation ceases, i.e. $\partial \Phi_b / \partial t = 0$, it leaves behind for the system the virial equilibrium state, $2 E_K/E_V=-1$, or equivalently, a hydrostatic equilibrium state, which is described by the Jeans equation \citep[see][]{galdyn08}
\begin{equation}
\label{eq:je}
\frac{\dd p}{\dd r} + \frac{2\beta p}{r} = -\rho \frac{\partial\Phi}{\partial r} = -\frac{G M \rho}{r^2},
\end{equation}
where the `pressure' $p$ is defined as $p \equiv \rho\sigma^2_r$, and $\sigma^2_r$ is the radial velocity dispersion. $\beta$ is the velocity anisotropy parameter, usually defined as $\beta \equiv 1-\sigma^2_t/\sigma^2_{r}$, where $\sigma^2_t$ is the one-dimensional tangential velocity dispersion, with $\sigma^2_t = \sigma^2_{\theta} = \sigma^2_{\phi}$.

The short-range residual effects of the gravitational forces between these particles are manifested in the following two forms of relaxation mechanisms. One is two-body relaxation, which arises from encounters of two bodies, and is only efficient locally, or at small scales of the system. At first, the system stays at its maximum radius with vanishing kinetic (or thermal) energy, and then is driven by violent relaxation to approach the virial equilibrium state. It is well known that, at this stage, two-body relaxation operates in a far longer time-scale than violent relaxation, so that it should actually be neglected when violent relaxation is functioning.

The other short-range relaxation mechanism operating at small scales in a self-gravitating system is gravitational Landau damping, which was first realized by \citet{lb62} and further studied by \citet{kandrup98}. A brief introduction to this short-range relaxation is the following. A density perturbation with wavelength shorter than the Jeans length, $\lambda_J$, will induce the Jeans waves; otherwise the perturbation will induce the Jeans instability. Such Jeans waves in self-gravitating systems, heavily damped by interactions with particles, are the counterparts of Langmuir waves in plasma. From the dispersion relation of the self-gravitating systems (equation~5.66 and fig.~5.2 of \citealt{galdyn08}), we can see that Landau damping is also very efficient locally. Hence, the two short-range relaxation mechanisms are responsible for the local equilibria of the system at small scales.

These relaxation mechanisms are picturized in Fig.~\ref{fig:fig1}(b). We can see that the three relaxation mechanisms, violent relaxation, two-body relaxation and Landau damping operate at different evolutionary stages and different spatial scales of the system: (1) two-body relaxation, which arises from interactions between particles themselves, is a short-range relaxation mechanism; (2) Landau damping, which is another short-range relaxation, arises from interactions between particles and the Jeans waves; (3) whereas if perturbations take place in large-scale regions in the early evolutionary stage, such large-scale perturbations will induce long-range violent relaxation, i.e. the large-range violent changing of the gravitational potential. We will see later the significance of the discrimination between these three types of relaxation mechanisms.

\subsection{Entropy stationary state}
\label{sec:stationary}

After violent relaxation terminates, the system stays in the hydrostatic equilibrium state, described by the Jeans equation, equation~(\ref{eq:je}). In this section, we have interests to investigate whether there are some extra physics resulting from violent relaxation besides the hydrostatic equilibrium.

As mentioned previously, there have been great disputes about the validity of the concept of entropy for self-gravitating systems for nearly five decades. Ignoring the controversy, we tentatively employ the entropy principle to explore the possible thermodynamic relations of self-gravitating dark matter haloes.

We choose the specific entropy in the phenomenological form as the following
\begin{equation}
\label{eq:specs}
s=\ln(p^{3/2} \rho^{-5/2}),
\end{equation}
which was first used by \citet{white87} for the isotropic self-bounded gas sphere, but here we assume it can also apply to the velocity-anisotropic case. This expression is the standard entropy form of an ideal gas in terms of $p$ and $\rho$, without the normalizing factor and an additive constant, and we borrow it for self-gravitating systems without any a priori reasons. With the specific entropy given, the total entropy of the system is
\begin{equation}
\label{eq:tots}
S_t = \int^{\infty}_0 4\pi r^2 \rho s \dd r = \int^{\infty}_0 4\pi r^2 \rho \ln(p^{3/2}\rho^{-5/2}) \dd r.
\end{equation}

As usual, the constraints are the total mass and energy conservation. The mass function $M(r)$ of a spherical gravitating system is
\begin{equation}
\label{eq:mf}
M(r) = \int^{r}_0 4\pi r'^2 \rho(r') \dd r',
\end{equation}
and the total mass of the system is
\begin{equation}
\label{eq:mt}
M_t = \int^{\infty}_0 4\pi r^2 \rho(r) \dd r.
\end{equation}
For the convenience of variational calculus, we set $y(r)=M(r)$. So the mass-conservation constraint is converted to satisfying the following boundary conditions:
\begin{equation}
\label{eq:bc}
y(r)|_{r=0}=0, {\hskip 1cm} y(r)|_{r=\infty}=M_t,
\end{equation}
and we replace $\rho(r)$ in the entropy functional with $y'(r)$, the first derivative of $y(r)$, as
\begin{equation}
\label{eq:rhod}
\rho(r)=\frac{1}{4\pi r^2}\frac{\dd M(r)}{\dd r} = \frac{y'(r)}{4\pi r^2}.
\end{equation}
The total binding energy of the system with spherical symmetry is
\begin{eqnarray}
\label{eq:ev}
E_V & = & -\int\frac{G \rho({\textbf{\emph r}})\rho({\textbf{\emph r}}')}{2 |{\textbf{\emph r}} - {\textbf{\emph r}}'|} \dd V \dd V' \nonumber \\
    & = & -4 {\pi} G \int^{\infty}_{0}M(r)\rho(r)r \dd r.
\end{eqnarray}
So by using the virial theorem, the total energy of the system is
\begin{equation}
\label{eq:et}
E_T=\frac{1}{2}E_V.
\end{equation}
The constrained entropy is
\begin{equation}
\label{eq:cs1}
S_{t,c}=S_t + \lambda E_t,
\end{equation}
where $\lambda$ is the Lagrangian multiplier. Substitute equations~(\ref{eq:tots}),
(\ref{eq:mf}), (\ref{eq:rhod}), and (\ref{eq:et}) into equation~(\ref{eq:cs1}), we have
\begin{eqnarray}
\label{eq:cs2}
S_{t,c}&=&\int^{\infty}_0 F(r, y, y') \dd r \nonumber \\
       &=&\int^{\infty}_0 \left[ y'\ln\left( p^{3/2} (\frac{y'}{4\pi r^2})^{-5/2} \right)-\lambda\frac{G y y'}{2 r}\right] \dd r.
\end{eqnarray}
We treat $p$ as a known function of $r$, and only calculate the first-order variation of $S_{t,c}$ with respect to $y$ and $y'$. We obtain the following Euler-Lagrangian equation from $\delta S_{t,c}=0$ as
\begin{eqnarray}
\label{eq:seq}
\frac{\dd s}{\dd r}&=&\frac{\dd}{\dd r}\ln\left( p^{3/2}(\frac{y'}{4\pi r^2})^{-5/2}\right) \nonumber \\
& = & \frac{\dd(\ln p^{3/2}\rho^{-5/2})}{\dd r} = -\lambda\frac{G M}{2 r^2}.
\end{eqnarray}
Since this equation comes from the first-order variational calculus, in what follows, we call equation~(\ref{eq:seq}) as the entropy stationary equation, or simply, the entropy equation. Additionally, from equation~(\ref{eq:specs}), it is interesting to notice that
\begin{equation}
s= \ln(p^{3/2}\rho^{-5/2}) = -\ln \frac{\rho}{\sigma^3_r}.
\end{equation}
That is, the so-called pseudo-phase-space density $\rho/\sigma^3_r$ seems to be related with the specific entropy $s$, which implies that the pseudo-phase-space density may really have a physical origin. This relation has been recognized by \citet[see equation~7.121]{galdyn08}. \citet{vass09} also suggested such a correlation between them that the phase-space density profile likely reflects the distribution of entropy, which dark matter acquires as it is accreted on to a growing halo.

With a simple dimensional analysis, we can write the Lagrangian multiplier $\lambda$ in equation~(\ref{eq:seq}) as
\begin{equation}
\label{eq:lambda}
\lambda=-\frac{\mu}{G \rho_s r^2_s},
\end{equation}
with $\mu$ being a positive dimensionless number, so that $\lambda$ is always negative. In the following, we can see that equations~(\ref{eq:je}) and (\ref{eq:seq}) have no solutions if $\lambda$ is positive. We can always set $\mu=1$ by choosing appropriate values for the characteristic density $\rho_s$ and scale $r_s$.

In this work, we use the radial pseudo-phase-space density $\rho/\sigma^3_r$ instead of the more general $\rho/\sigma^3$ to formulate the specific entropy, which by no means indicates that the former is physically superior to the latter. In fact, we have tried many combinations of $\sigma$, but all failed to yield reasonable results. For example, take the form of $\sigma$ as $\sigma^3 = \sigma_r \sigma_{\theta} \sigma_{\phi}=\sigma^3_r (1 - \beta)$, and perform the same variational calculus as for equation~(\ref{eq:seq}), we get the equation
\begin{equation}
\label{eq:seq2}
\frac{\dd\ln(p^{3/2}\rho^{-5/2})}{\dd\ln r}-\frac{1}{(1-\beta)} \frac{\dd\beta} {\dd\ln r} = -\lambda \frac{G M}{2 r},
\end{equation}
or with $\sigma^3=(\sigma^2_r + \sigma^2_{\theta} + \sigma^2_{\phi})^{3/2} = \sigma^3_r (3 - 2\beta)^{3/2}$, we derive the following result:
\begin{equation}
\label{eq:seq3}
\frac{\dd\ln(p^{3/2}\rho^{-5/2})}{\dd\ln r}-\frac{3}{(3 - 2\beta)} \frac{\dd\beta} {\dd\ln r} = -\lambda \frac{G M}{2 r}.
\end{equation}
Compared with the entropy stationary equation~(\ref{eq:seq}), it can be seen that there appears an additional term, $ -\dd{\beta} / \dd\ln r (1-\beta)$, or $ -3 \dd{\beta} / \dd\ln r (3 - 2\beta)$ in the above two equations. From fig.~9 of Navarro et al. (2008), we can see that $\beta$ changes sharply at the outskirts of the dark matter halo, so the absolute value of $\dd\beta/\dd\ln r$ at the outer boundary is huge. Such an additional term containing $\beta$ will greatly affect the behaviour of the solution of equations~(\ref{eq:seq2}) or (\ref{eq:seq3}), so that we cannot even get a convergent solution.

Moreover, there is also another problem that we cannot perform the variational calculus with respect to the pressure $p$, since such a resultant equation is meaningless and unacceptable, so in the current work we have to treat $p$ as a known function. As a consequence, such a treatment is inconsistent, and hence at this moment we can obtain only one equation by using the entropy principle. Because of this inconsistency, our results are incomplete.

These problems indicate the great complexity of the statistical physics of self-gravitating systems, and imply that, either the entropy form may be even more complicated than the current forms we have already considered, or we may still lack some crucial physics, such as some extra macroscopic constraints besides the mass and energy conservation. We believe these problems should be the targets of future investigations.

\section{Results}
\label{sec:result}

Up to now, for the three independent variables $p$, $\rho$ and $\beta$, we just have two equations, the Jeans equation (\ref{eq:je}), and the entropy equation (\ref{eq:seq}). In order to obtain definite solutions of them, we have to provide an extra equation to close the equation system. This can be done by assigning one of the three variables with a known function. In the following, we explore the solutions in two cases, respectively: (1) given the anisotropy parameter $\beta$, and (2) given the density profile $\rho$.

\subsection{Given anisotropy parameter $\beta$}
\label{ssec:beta}

\begin{figure}
\includegraphics[width=\columnwidth]{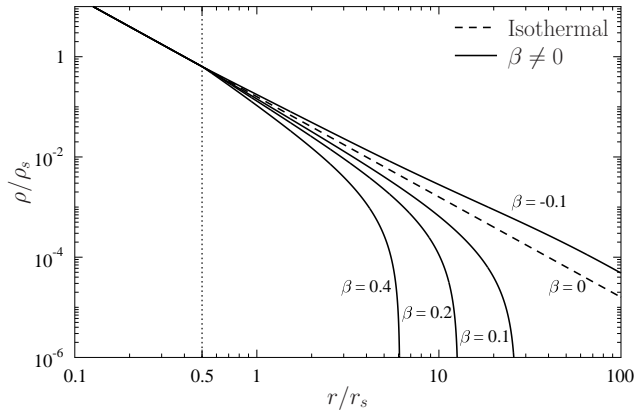}
\caption{Density functions as the solution of equations~(\ref{eq:je}) and (\ref{eq:seq}), with $\beta$ being a given function of $r$ in equation~(\ref{eq:beta}). The fiducial solution of $\beta=0$ is indicated as `Isothermal' in the figure; others are shown as $\beta \neq 0$.}
\label{fig:fig2}
\end{figure}

With the anisotropy parameter $\beta$ given, the density profiles can be solved, and these solutions are shown in Fig.~\ref{fig:fig2}. For simplicity, we specify $\beta$ as a step function of $r$ as
\begin{eqnarray}
\label{eq:beta}
\beta(r) = \left\{
\begin{array}{ll}
  ~~0, & \mbox{$r \leqslant C r_s$},  \\
 \beta_0, & \mbox{$r > C r_s$},
\end{array}\right.
\end{eqnarray}
in which $C$ is arbitrarily specified as, say $0.5$, and the values of the constant $\beta_0$ are indicated in Fig.~\ref{fig:fig2}. First, we concentrate on the case of $\beta=0$. Here, with the Lagrangian multiplier $\lambda$ chosen to be negative, the two equations have neat analytic solutions as $p\propto r^{-2}$ and $\rho \propto r^{-2}$, which are exactly the isothermal solutions. It is well known that in such a solution, the system has infinite mass, energy and extent, which is similar to \citet{lb67}'s result, and so is not acceptable.

Such an unconfined density configuration of self-gravitating systems once misguided \citet{lb67} and \citet{shu78} to believe that violent relaxation is not a complete relaxation mechanism, and so introduce the quest for a theory of incomplete relaxation \citep[see also][]{madsen87}.

However, we can see that cases with $\beta > 0$ are different from the isothermal solution. These solutions gradually deviate downwards from the isothermal solution and descend more and more rapidly with increasing radius. The larger the value of $\beta$, the more decline of the solution, and the density functions drop down sharply at some outer radii. It is obvious to see that $\beta$ is of great significance to make the solutions have finite mass, energy and spatial extent, and therefore, with $\beta$ involved, it is not necessary to introduce a phase-space cut-off, or incomplete relaxation mechanisms \citep{lb67, shu78}.

\subsection{Given density profile $\rho$}
\label{ssec:rho}

Now let us discuss the other case, i.e. solving equations~(\ref{eq:je}) and (\ref{eq:seq}) by specifying a dark halo density profile $\rho(r)$. Here, we use the \citet{einasto65} profile \citep[also see][]{navarro08},
\begin{equation}
\label{eq:einasto}
\ln(\rho(r)/\rho_s) = -\frac{2}{\alpha}[(r/r_s)^{\alpha}-1],
\end{equation}
with the shape parameter $\alpha=0.170$, the same as the one taken by \citet{navarro08} for their Aq-A halo simulation series.

\begin{figure}
\centerline{\includegraphics[width=\columnwidth]{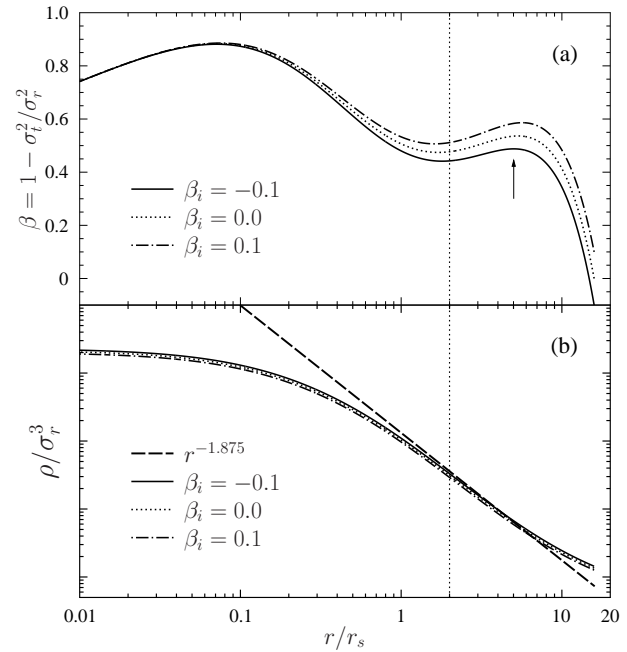}}
\caption{(a) $\beta(r)$ as one of the solutions of equations~(\ref{eq:je}) and (\ref{eq:seq}), with the density profile given as the Einasto profile in equation~(\ref{eq:einasto}). We numerically integrate the two equations towards the inner of the dark matter halo, from the outer boundary at $r = 16 r_s$, where the logarithmic slope of the Einasto profile is $\dd\ln\rho / \dd\ln r \approx -3.2$. The initial values of $\beta$ are shown in the figure. The arrow indicates the position of the local peak, at about $r \approx 5 r_s$. (b) The other solution of equations~(\ref{eq:je}) and (\ref{eq:seq}), shown as the radial pseudo-phase-space density profile $\rho/\sigma^3_r$ as functions of radius $r$. The vertical scale is unnormalized. }
\label{fig:fig3}
\end{figure}

We solve the two equations from the outer boundary towards the inner of the dark matter halo. The two solutions are shown as $\beta(r)$ and the radial pseudo-phase-space density $\rho/\sigma^3_r$ in Fig.~\ref{fig:fig3}. Compared with the corresponding simulation results in \citet[see figs~9 and 12]{navarro08}, we can see that our solutions in the inner part of the dark matter halo are completely unacceptable. We ascribe this disagreement between our theoretical predictions and the simulation results to the phenomenon called gravitational degeneracy \citep{chavanis98}, which is not included in this paper. We briefly introduce this effect in the following.

\citet{lb67} derived a distribution of self-gravitating stellar systems that is analogous to the Fermi-Dirac distribution. In his work, the gravitating particles are grouped into, and represented by the coarse-grained, constant-density phase-space elements, which have fixed and finite phase volume, and cannot overlap to each other. It is this Fermi-Dirac-type distribution that leads to the existence of the gravitational degeneracy in the inner region of self-gravitating systems, where is much denser than the other parts. Such a gravitational degeneracy provides an exclusive effect against the gravitational collapse to avoid the gravothermal catastrophe \citep{lb68}. It should not be confused that the gravitational degeneracy results not from Pauli exclusion principle of quantum mechanics, but from the conservation of the overall phase-space volume of these phase elements. In strong degenerate region, i.e. the inner part of the dark halo, the close encounters should not be ignored so that the specific entropy, equation~(\ref{eq:specs}), which is just suitable for ideal gas composed of point-like particles, is not proper. It is valid only in the non-degenerate limit, i.e. the outer low-density region of the dark halo.

If the degenerate effect is included, the proper specific entropy may be better generalized to, say, the \citet{white87}'s form
\begin{equation}
\label{eq:dess}
s = -\ln F-(\frac{\eta}{F}-1) \ln(\eta - F),
\end{equation}
where $F=\rho/\sigma^3_r$ is the pseudo-phase-space density, and $\eta$ is the maximum phase-space density, which embodies an exclusive effect that prevents different phase elements from overlapping to each other in the final state. In the current work, however, we restrict ourselves not to include this complexity of gravitational degeneracy.

According to the behaviour of the solutions, we draw a vertical dotted line at $r = 2 r_s$ in Fig.~\ref{fig:fig3}, and an arrow pointing to the local peak at $r \approx 5r_s$ in the upper panel, to roughly separate the halo into the innermost strong-degenerate (with the radius $r<2r_s$), the middle weak-degenerate ($2r_s < r < 5r_s$) and the outmost non-degenerate ($5r_s < r < 16r_s$) region. As mentioned above, due to the ignoring of gravitational degeneracy, our theoretical predictions in the inner strong-degenerate region of the halo are not reliable at all. So in the following we just concentrate on the solutions in the outer regions.

The velocity-anisotropy parameter for dark matter haloes that is greater than for baryons has been observationally conformed \citep{hansen07, host09}, whereas the latest high-resolution simulations revealed much more details that are absent in the observations. \citet{navarro08} and \citet{ludlow10} found the non-monotonic radial dependence of $\beta$, i.e. haloes are nearly isotropic near the center, radially biased to the maximum at some radius and approximately isotropic again in the outskirts, which is against the earlier result of \citet{hansen06}, who found that the anisotropy parameter just increases monotonically outwards from the inner part of the dark matter halo.

From Fig.~\ref{fig:fig3}, we see that the predicted $\beta$ agrees quantitatively well with the corresponding simulation results of \cite{navarro08} in the non-degenerate region. In particular, it is worth emphasizing that our theoretical predictions completely reproduce the non-monotonic radial dependence of $\beta$ outside the strong-degenerate kernel. Besides, our predictions are also in qualitative agreement with the simulation results in the middle weak-degenerate region of the dark halo.

Similar to the case of $\beta$, the predicted pseudo-phase-space densities $\rho / \sigma^3_r$ in the inner strong-degenerate kernel also greatly deviate from the simulated results, for the same reason as that for $\beta$, i.e. ignoring of the gravitational degeneracy. In the outer regions with $r>2r_s$, however, the predictions, similar to the simulation results, can also be well fitted by $\rho / \sigma^3_r \sim r^{-1.875}$ -- the power law predicted by \citet{bertschinger85}'s similarity solution for infall onto a point mass in an otherwise unperturbed Einstein-de Sitter universe -- except for an upward deviation in the halo outskirts. We are delighted to see that our result is in good agreement with the finding by \citet{ludlow10}, that such a curve-up deviation also exists in their simulations. This significant upward deviation from the simple power law of $\rho / \sigma^3_r \sim r^{-1.875}$ indicates that the great complexity exists in equilibrated dark matter haloes \citep*{schmidt08}.

Thus, we have seen that, compared with the simulations, our theoretical results provide good agreements in the outer non-degenerate region of the dark halo, and also acceptable agreements in the weak-degenerate region, respectively. We conclude that these agreements between our theoretical predictions and simulation results sufficiently indicate that both the concept of entropy and the entropy principle should be still valid for long-range self-gravitating systems.

\begin{figure}
\centerline{\includegraphics[width=\columnwidth]{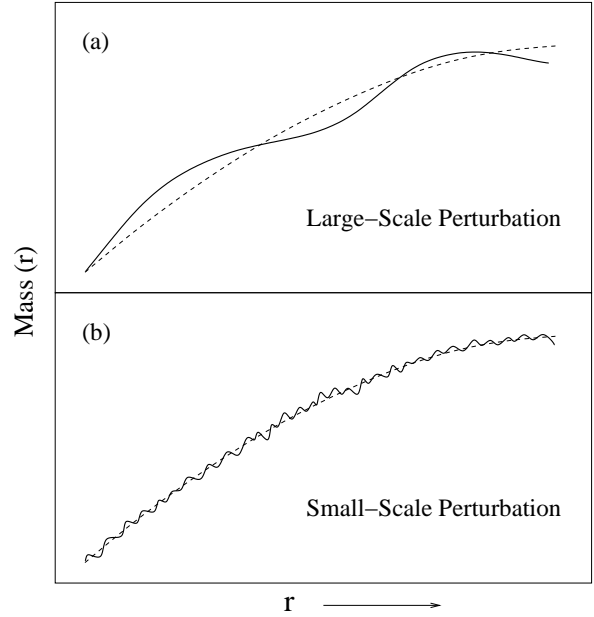}}
\caption{Two modes of perturbation. The dashed lines in both panels indicate the equilibrium configurations. (a) Large-scale mass perturbation, i.e. $|A|(\delta y)^2 \gg |B|(\delta y')^2$; (b) small-scale density perturbation/fluctuation, i.e. $|A|(\delta y)^2 \ll |B|(\delta y')^2$, where $A$ and $B$ are defined in equation~(\ref{eq:ab}).}
\label{fig:fig4}
\end{figure}

\section{Generalization of the Second Law of Thermodynamics}
\label{sec:gen2nd}

\subsection{Second variation}
\label{sec:var2}

We would like to inspect whether the solution of the first-order variational calculus, i.e. equation~(\ref{eq:seq}), is a maximum, minimum or any other types of solutions. For this sake, we calculate the second-order variation. By straightforward substitution, the second-order variation of the constrained entropy of equation~(\ref{eq:cs2}) is
\begin{eqnarray}
\label{eq:2ndvar}
\delta^2 S_{t,c} & = & \frac{1}{2}\int^{\infty}_0 [A(\delta y)^2 + B(\delta y')^2] \dd r \nonumber \\
& = & \frac{1}{2} \int^{\infty}_0 [A(\delta M)^2 + B(4\pi r^2 \delta \rho)^2 ] \dd r,
\end{eqnarray}
where
\begin{eqnarray}
\label{eq:ab}
A & = & \frac{\partial^2 F}{\partial y^2} - \frac{\dd}{\dd r}\left(\frac{\partial^2 F} {\partial y\partial y'}\right) = -\lambda\frac{G}{2 r^2}, \nonumber \\
B & = & \frac{\partial^2 F}{\partial y'^2} = -\frac{5}{2 y'} = -\frac{5} {8\pi
r^2\rho}.
\end{eqnarray}
For the convenience of the following analysis, we call the first term in the above integral, $A(\delta M)^2$, as {\em mass perturbation}, and the second term, $B(4\pi r^2 \delta \rho)^2$, as {\em density perturbation}. According to the result of the first-order variational calculus, we know $\lambda<0$, so we can immediately see that $A>0$, and $B<0$. This result indicates that the entropy stationary solution, equation~(\ref{eq:seq}), is neither a maximum nor a minimum, but a saddle-point solution of $\delta S_{t,c}=0$. So at first sight, it seems that the solution might be unstable, and the entropy form of equation~(\ref{eq:specs}) may not be proper for self-gravitating systems.

In fact, this apparent catastrophe contains a profound new physics that we may have never realized before. In what follows, we will analyze it in detail.

\begin{figure*}
\centerline{\includegraphics[width=1.5\columnwidth]{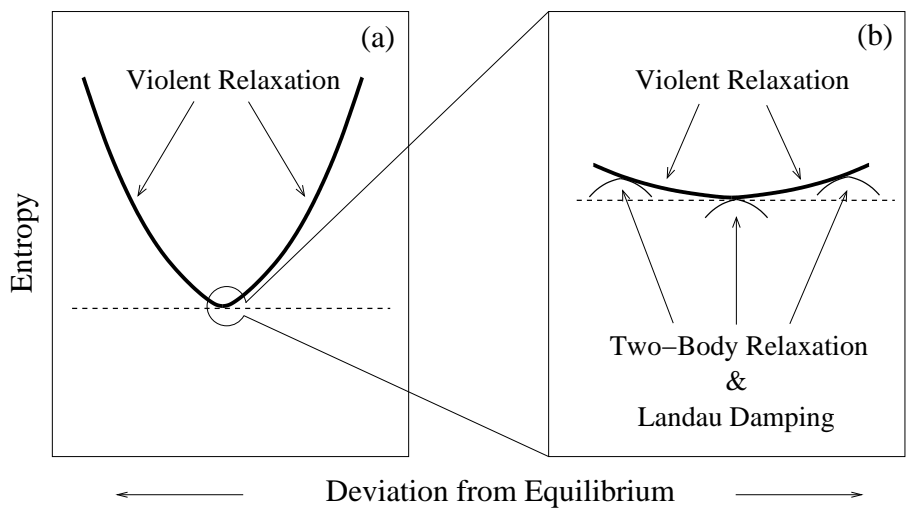}}
\caption{Illustration of the effects of the three relaxation mechanisms. The horizontal dashed lines in both panels indicate the entropy in equilibrium states. (a) If the system deviates far beyond its equilibrium state, it will be driven by violent relaxation back to its equilibrium again. (b) Magnification of the local near-equilibrium state in panel (a). Under long-range violent relaxation, the system approaches its global minimum entropy state, whereas short-range two-body relaxation and Landau damping collectively operate to maintain this very entropy as the local maximum for every and any small part of the system.}
\label{fig:fig5}
\end{figure*}

\subsection{The complementary second law of thermodynamics for self-gravitating systems}
\label{sec:rev2nd}

First, we assume that $|A|(\delta y)^2 \gg |B|(\delta y')^2$, or equivalently,
\begin{equation}
\label{eq:lrelax}
-\lambda \frac{G}{2 r^2}(\delta M)^2 \gg \frac{10 \pi r^2}{\rho}(\delta\rho)^2.
\end{equation}
Here, the mass perturbation is predominant, whereas the density perturbation, i.e. the variation of the first derivative of mass function, is negligible. As a result, such a mass perturbation must be large-scale, which is shown in Fig.~\ref{fig:fig4}(a). One may ask what physical processes can cause such kind of mass perturbation. A rough analysis shows that the dynamical processes, such as gravitational collapse or the system's radial large-range oscillations with large amplitude, can all lead to the mass perturbations, which must induce the violent changing of the long-range, background gravitational potential with respect to time, i.e. $\partial \Phi_b / \partial t \neq 0$. This is exactly the feature of violent relaxation.

So from equation~(\ref{eq:2ndvar}), with $A>0$ and without small-scale density perturbation, we see that the entropy of the isolated self-gravitating systems, driven by violent relaxation, will always {\em automatically decrease} to the minimum until violent relaxation terminates. Furthermore, we notice that the mass perturbation, $-\lambda G (\delta M)^2 /2 r^2$, does not contain any localized information, say, some function of the density $\rho(r)$, hence we can further identify this equilibrium entropy as the global minimum entropy.

Our result is consistent with Binney's argument, which was designed to argue against the validity of Boltzmann-Gibbs entropy as well as the maximum entropy principle for self-gravitating systems. This argument briefly states as follows. The total mass and total energy of the system are assumed to be $M_T$ and $E_T$ (negative), respectively. A large fraction of the mass and the energy, say $M_T-\mu$ and $E_T-\epsilon$, are concentrated in a bound static core, whereas the remaining mass $\mu$ and thermal energy $\epsilon$ are deposited in a diffuse outer envelope. As $\epsilon\rightarrow 0$ the envelope becomes larger and larger in radius, but its density becomes more and more tenuous, such that the associated entropy will diverge to infinity \citep{galdyn08}. So the fact that entropy is unbounded above is consistent with our finding.

Besides Binney's argument, \citet{ant62} also proved that global maximum entropy states do not exist for self-gravitating systems (see also appendix III of \citealt{lb68}), which is also a support to our conclusion.

\begin{figure*}
\centerline{\includegraphics[width=1.5\columnwidth]{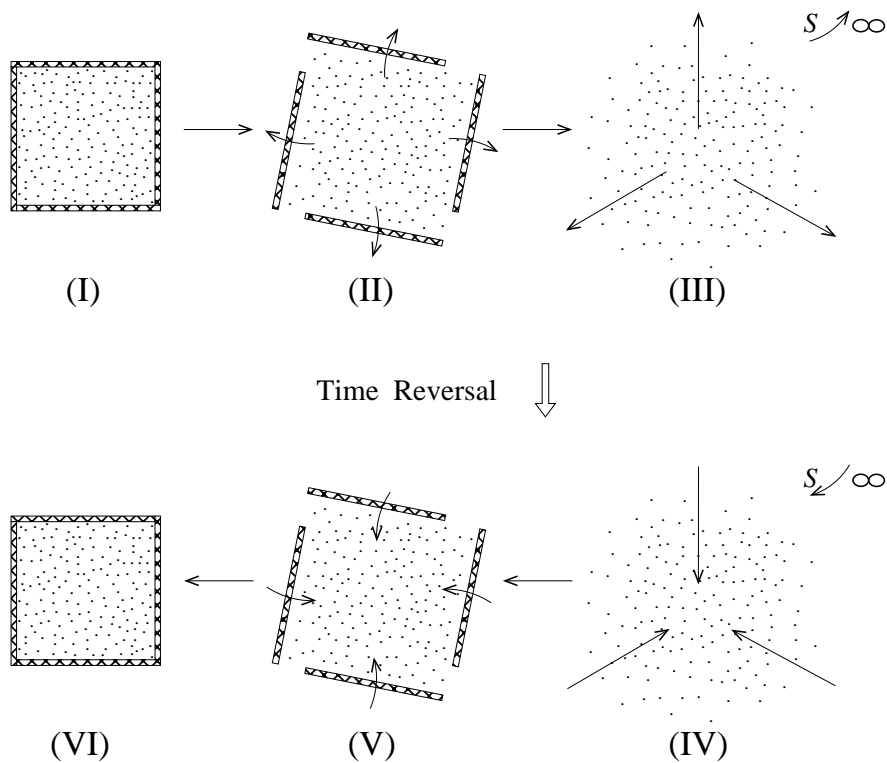}}
\caption{A thought experiment helps to demonstrate the global minimum--local maximum entropy states of self-gravitating systems. Arrows indicate the time direction. From (I) to (III) are the free expansion of ideal gas; from (IV) to (VI) are the time reversal of the above procedure.}
\label{fig:fig6}
\end{figure*}

Now we turn to the analysis of density perturbation. Here, we have $|A|(\delta y)^2 \ll |B|(\delta y')^2$, or equivalently,
\begin{equation}
\label{eq:lrelax}
-\lambda \frac{G}{2 r^2}(\delta M)^2 \ll \frac{10 \pi r^2}{\rho}(\delta\rho)^2.
\end{equation}
In the context of density perturbation, the mass perturbation is negligible, which implies $\partial \Phi_b/\partial t \simeq 0$, i.e. there is no violent relaxation. So we can see from Fig.~\ref{fig:fig4}(b) that the density perturbation is just small-scale perturbation. Such small-scale density perturbations will induce only short-range relaxations, i.e. $\partial \delta \Phi / \partial t \neq 0$, such as two-body relaxation and Landau damping. With $B<0$ and without mass perturbation, we can see from equation~(\ref{eq:2ndvar}) that the entropy is the maximum for the system under short-range relaxations. Moreover, because of the small-scale nature of the density perturbation, we can further identify this equilibrium entropy as the local maximum entropy, not for the whole entity, but just localized for every and any small part, or roughly, coarse-grain, of the system.

We further illustrate the differences of the long- and short-range relaxation mechanisms in Fig.~\ref{fig:fig5}. When the system has experienced a large-scale mass perturbation and departures greatly from the equilibrium state, long-range violent relaxation will be excited and drive the system back to the equilibrium state, i.e. the minimum entropy state. Nevertheless, if violent relaxation nearly or completely terminates, then short-range two-body relaxation/Landau damping operate locally to maintain the local maximum entropy state for every/any small part of the system. In fact, in the usual thermodynamics with no long-range relaxation mechanisms, if the system is extremely large and unconfined, such as the atmosphere of the Earth, it is impossible to acquire a global, but just local thermodynamic equilibria, for the system. So we can see that things are indeed complicated for the self-gravitating system, but our results do not conflict with the standard second law of thermodynamics at all, since the latter is just valid for the systems dominated by short-range relaxation mechanisms. We here just generalize the second law to accommodate the long-range, self-gravitating systems.

To summarize, we re-express the second law of thermodynamics for self-gravitating systems in the seemingly paradoxical, but actually complementary way as follows: for an isolated self-gravitating system, if long-range relaxation (i.e. violent relaxation) is predominant, the entropy of the system never increases; whereas if short-range relaxations (i.e. two-body relaxation together with Landau damping) are predominant, the entropy never decreases. An additional explanation for this revision of the second law of thermodynamics states that the equilibrium states of self-gravitating systems are the global minimum state for the whole system under long-range relaxation, but the local maximum entropy state for every and any small part of the system under short-range relaxations. It is only for the long-range self-gravitating systems that there exists such a minimum-maximum entropy duality, and the corresponding complementarity for the second law of thermodynamics. For the usual thermodynamic systems, there is no distinction between long-range and short-range relaxations, so there exists no such a complementarity for the standard second law of thermodynamics.

As addressed in the Introduction, some earlier authors \citep[e.g.][]{sridhar87, soker96} have already realized that the $H$-function, or entropy of self-gravitating systems, may not necessarily increase during the violent relaxation process. \citet{soker96} even distinguished between two phases of violent relaxation, in which the $H$-function may decrease with time in the first phase, i.e. the `vigorous phase' they called, whereas in the second phase, i.e. the `calm phase' they called, it is always a non-decreasing function of time. If we tentatively regard Landau damping as the second phase of violent relaxation as they referred to, one can notice that their finding is very close to our result, which is a great support to the complementarity we found in this work. However, compared with our distinction between the three relaxation mechanisms, their two-phase classification seems slightly vague and inaccurate, and they did not relate these two phases to the long- and short-range relaxations, either. Most of all, it seems that they did not realize the new physics, i.e. the complementary second law of thermodynamics, underlying such a decreasing/incresing property of the $H$-function.

We propose a thought experiment in Fig.~\ref{fig:fig6}, which may help to understand this generalized, or complementary, second law of thermodynamics. The processes are the following. (I) An ideal gas is confined within an adiabatic box, which resides in the vacuum. The short-range relaxations, namely, collisions of the particles between themselves and with the walls of the box, maintain the ideal gas in its thermal equilibrium, i.e. the maximum entropy state. (II) Suddenly remove the walls, and the gas begins to expand outwards. (III) This is just the familiar free-expansion process of an ideal gas, in which the gas will expand all the way without ending, and we know that the entropy will increase to infinity along with the expansion. Now we time reverse the above processes from (I) to (III), and then the time-reversal processes proceed in this way: (IV) the gas moves inwards to concentrate to a small spatial region, pulled by some fictitious `long-range attractive force', and of course the entropy of the gas decreases with this process. (V) The ideal gas is suddenly enclosed by adiabatic walls, and then the `long-range relaxation' stops, leaving behind a mechanical equilibrium of the ideal gas. The entropy of the system also stops decreasing, so we may say the system attains its minimum entropy state. (VI) The equilibrated ideal gas implemented by the `long-range relaxation' is then maintained by the same short-range relaxations as in stage (I), so that the entropy is the maximum under the short-range relaxations. We can see that, except for the sudden establishing of the potential well, i.e. the adiabatic walls, this time-reversal procedure greatly resembles the gravitational collapsing process under the action of gravitational force. This thought experiment helps to understand the seemingly strange behavior of the minimum -- maximum entropy states of self-gravitating systems.

\section{Summary and conclusions}
\label{sec:cons}

The statistical mechanics of self-gravitating systems has long been believed as an inviable discipline. The controversy mostly focuses on whether the concept of entropy is still valid for long-range interaction systems, such as self-gravitating systems. In this work, we perform a preliminary study to this age-old problem. We summarize our findings in the following.

When self-gravitating systems attain their equilibrium states, the systems generally possess some geometrical symmetry, such as a spherical shape. Under this condition, it is desirable to approximate the self-gravitating systems by the mean-field model. That is, the long-range part of gravitational forces between the particles is collectively manifested as a smooth background force field, with the short-range residual effects determined by nearby particles. Such a separation of long- and short-range gravitational force breaks the long-range coupling of a particle with the others at all different range of scales, and definitely renders the problem more manageable.

Entropy is a state function that is introduced to describe the degree of order or chaos of a thermodynamic system. This concept, however, is of great significance in statistical mechanics, not just because it is a thermodynamic state function, but also because it constitutes a variational principle called the entropy principle (with the obsolete name `principle of maximum entropy'), from which the equations or relations of thermodynamic quantities can be derived. It plays the same role for statistical mechanics as the one played by the principle of least action for fundamental dynamics. Without an appropriate definition of entropy of the thermodynamic system under consideration, it is impossible to establish the statistical mechanics for this system, as has been exactly the case of self-gravitating systems for the past several decades. In this work, we tentatively employ the phenomenological entropy form of ideal gas, first used by \citet{white87}, to explore the possible thermodynamic relations of the equilibrium states of self-gravitating systems.

By calculating the first-order variation with respect to the density function $\rho$, subject to the constraints of mass and energy conservation, we derive an entropy stationary equation. However, we acknowledge that our treatment is somewhat deficient in completeness or consistency, in that we have to treat the pressure $p$ as a known function, and cannot calculate the variation with respect to it. As a result, we can derive only one equation, i.e. the entropy stationary equation. Moreover, we have also tried some other entropy forms other than the \citet{white87} form, but all failed to yield any reasonable results. These problems imply that either the entropy form may be even more complicated than the current forms we have already considered, or we may still lack some crucial physics, such as some extra macroscopic constraints besides the mass and energy conservation. However, despite this incompleteness/inconsistency, when incorporated with the Jeans equation, we do find some interesting results with this entropy equation.

In the case of given anisotropy parameter $\beta$, we solve the entropy stationary equation and the Jeans equation to obtain $p$ and $\rho$. If $\beta = 0$, the two equations possess the isothermal solutions. However, solutions with $\beta > 0$ deviate downwards from the isothermal solution, descend more and more rapidly with increasing radius, and drop down sharply at some outer radii. We conclude that $\beta$ is of great significance to make the solutions have finite mass, energy and spatial extent, and hence it is not necessary to introduce a phase-space cut-off, or incomplete relaxation mechanisms. According to the behaviour of the solutions, we separate the halo into the inner strong-degenerate ($r < 2 r_s$), the middle weak-degenerate ($2 r_s < r < 5 r_s$) and the outer non-degenerate ($5 r_s < r < 16 r_s$) region. In the case of given density profile $\rho$, say the Einasto profile, we derive the solutions of $\beta$ and $\rho/\sigma^3_r$, which are in good agreements with the simulation results of \citet{navarro08} in the outer non-degenerate region of the dark halo, and also in rough agreements with the simulations in the middle weak-degenerate region. In particular, it is worth emphasizing that our theory predicts an upward deviation of $\rho/\sigma^3_r$ from the power law, $\rho/\sigma^3_r \sim r^{-1.875}$, at the halo outskirts, which is consistent with the finding by \citet{ludlow10}, that such a curve-up deviation also exists in their simulations. However, we cannot explain the simulation results in the inner strong-degenerate region, since we have not considered the effect of gravitational degeneracy in the current theory.

By calculating the second-order variation of the entropy functional, we find that the stationary solution of the first-order variation is neither a maximum nor a minimum, but a saddle-point solution. We argue that such a saddle-point solution by no means indicates a failure of the theory, but rather suggests a new physics that we may have never realized before. Inspired by this saddle-point solution, we distinguish between two types of perturbations in self-gravitating systems, namely, the large-scale mass perturbation and the small-scale density perturbation, which correspond to long-range violent relaxation and short-range two-body relaxation/Landau damping, respectively. These two types of relaxation mechanisms operate in different fashions. For an isolated self-gravitating system, the entropy can automatically decrease under long-range violent relaxation, until it terminates and the system attains the equilibrium state with the global minimum entropy. This result is consistent with Antonov's proof, or Binney's argument that there are no global maximum entropy states for self-gravitating systems. However, for every and any small part of the system in its equilibrium state, the entropy is actually the local maximum under short-range relaxations, which is not in conflict at all with the standard second law of thermodynamics. This minimum-maximum entropy duality constitutes the complementary second law of thermodynamics for long-range self-gravitating systems. \citet{soker96} found the decreasing/increasing property of the general $H$-functions in the two-phased violent relaxation, which can be regarded as a support to our finding in this work.

Although we have not proposed a complete entropy form, it is extremely intriguing that this simple, or somewhat crude, entropy form of ideal gas, by \citet{white87}, can lead to such a profound new physics. Nevertheless, our study has definitely demonstrated that the concept of entropy is still valid for self-gravitating systems. However, the standard second law of thermodynamics should be revised, and the term `principle of maximum entropy' should be abandoned, and should be simply replaced by `entropy principle'. The validity of entropy for self-gravitating systems means that the statistical mechanics and the relevant thermodynamic relations can also, in principle, be established for such a long-range interaction system. We believe that our findings, especially the complementary second law of thermodynamics, are interesting, and may offer important clues to the development of statistical mechanics for self-gravitating systems as well as other long-range interaction systems.

\section*{Acknowledgements}

PH is very grateful for the comments and suggestions by Professor Simon D. M. White, Drs. S. Hansen, and A. Graham. This work is supported by the Chinese Academy of Sciences under Grant No. KJCX3-SYW-N2, also in part by the Project of Knowledge Innovation Program (PKIP) of Chinese Academy of Sciences under Grant No. KJCX2.YW.W10 and by National Basic Research Program of China, No.~2010CB832805.


\label{lastpage}
\end{document}